\documentclass[a4paper]{article}
\usepackage{spconf,epsfig}
\usepackage{multicol}
\usepackage{graphicx}
\usepackage{subfigure}
\usepackage{amsmath}
\usepackage{amsfonts}
\usepackage{color}
\usepackage{epstopdf}
\usepackage{algorithm}
\usepackage{algorithmic}
\usepackage{multirow}
\usepackage{verbatim}
\usepackage{setspace}
\newcommand{\be}{\begin{equation}}
\newcommand{\ee}{\end{equation}}

\title{Designing dedicated data compression for physics experiments\\ within FPGA already used for data acquisition}
\name{Jarek Duda$^{\star}$ \qquad Grzegorz Korcyl$^{\dagger}$\qquad }
\address{$^{\star}$ {Faculty of Mathematics and Computer Science, Jagiellonian University, Krakow, Poland}\\
   $^{\dagger}$ {Faculty of Physics, Astronomy and Applied Computer Science, Jagiellonian University, Krakow, Poland}}

\begin{document}
\ninept
\maketitle

\begin{abstract}
Physics experiments produce enormous amount of raw data, counted in petabytes per day. Hence, there is large effort to reduce this amount, mainly by using some filters. The situation can be improved by additionally applying some data compression techniques: removing redundancy and optimally encoding the actual information. Preferably, both filtering and data compression should fit in FPGA already used for data acquisition - reducing requirements of both data storage and networking architecture.

We will briefly explain and discuss some basic techniques, for a better focus applied to design a dedicated data compression system basing on a sample data from a prototype of a tracking detector: 10000 events for 48 channels. We will focus on the time data here, which after neglecting the headers and applying data filtering, requires on average $\approx$1170 bits/event using the current coding. Encoding relative times (differences) and grouping data by channels, reduces this number to $\approx$ 798 bits/channel, still using fixed length coding: a fixed number of bits used for a given value. Using variable length Huffman coding to encode numbers of digital pulses for a channel and the most significant bits of values (simple binning) reduces further this number to $\approx$ 552bits/event. Using adaptive binning: denser for frequent values, and an accurate entropy coder we get further down to $\approx$ 455 bits/event - this option can easily fit unused resources of FPGA currently used for data acquisition. Finally, using separate probability distributions for different channels, what could be done by a software compressor, leads to $\approx$ 437bits/event, what is 2.67 times less than the original 1170 bits/event.
\end{abstract}

\section{introduction}
\nointerlineskip
Continuous development of measurement techniques, precision and readout rates, results in increasing data volume generated by readout systems in modern physics experiments. Currently used data reduction mechanisms rely on fast, on-line data analysis and filtering performed by hardware modules equipped with FPGA devices. Those methods implement low-level algorithms, which usually are limited by the architecture of readout systems (e.g. only separate parts of the entire detector can be analyzed on one device) and the nature of FPGAs (e.g. basic arithmetic functions, limited data buffering capabilities). Therefore, such filters select interesting data based on partial information and reject data considered as not interesting, which under more extensive analysis could turn out to be valuable. The great challenge in the design of acquisition system is the right balance between the final amount of generated data and its physical quality.

Large data volumes require advanced networking infrastructure and expensive storage space, as the data from detector runs is supposed to be kept for decades. Hence introduction of advanced, adaptive data structures and compression algorithms can help reducing both: costs of developing and running experiments and the amount of valuable data rejected by real-time filtering mechanisms.\\

We will discuss applying general techniques to design a dedicated data compression system. For a better focus, we will do it basing on a sample of timing data from a prototype of a tracking detector: 10000 events from 48 channels, having in mind required scaling to a larger number of channels. While a software archiver would be also useful, the primary motivation is putting data compression into FPGA already used for data acquisition (employing its unused resources). In this way we will be able not only to reduce  storage media usage, but also lower transmission requirements and improve reading time, as decompression ($\approx$ 500MB/s/core) is usually much faster than reading from a data medium (50-120 MB/s for HDD).

\begin{figure}[t!]        
    \centering
        \includegraphics[width=8.5cm]{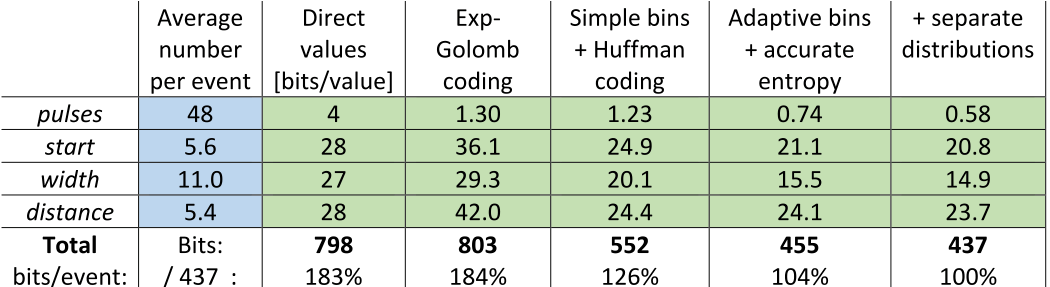}\
\begin{center}
        \caption{The summary of discussed approaches - while the current coding uses on average 1170 bits/event, the best discussed one requires 437 bits/event. The cost consists of storing 4 types of values: \emph{distance} is time difference between successive activations, \emph{width} is time difference between rising and falling time for a given activation, \emph{start} is the time of first activation (is zero once per event), \emph{pulses} is the number of digital pulses for a given channel. }
\end{center}
\label{table}
\end{figure}

Every event corresponds to a particle passing through several channels of the detector and hitting the reference detector. The electronics convert analog impulses from detector channels and produces digital signals with a certain width. Time-to-Digital Converter measures the time of rising and falling edge of such signal and returns a numeric value representing absolute time. We assume that faulty measurements (e.g. one edge missing or corrupted value) are already filtered in FPGA and therefore we only have to encode successive pairs of times (rising and falling).

For simplicity and clearance we will omit the general data packet structure (headers) and focus only on essential data coming from time-measurement devices: storing times within an event. This data uses on average 1170bits/event in the current data format, the columns of Fig. \ref{table} contain comparison with successive optimizations that we will discuss here. The one before last (455 bits/event) is suitable to fit FPGA, the last one can be implemented in a software compressor.\\

This analysis assumes that we encode only meaningful data which will be used in the data processing phase. This means shifting filtering close to digitization, favorably to FPGA, what results in saving of usage of both data storage and transmission lines. Counterargument against such approach is that unfiltered data can be used for the diagnostic purposes. To resolve this issue, there can be used two modes: the main run data saving mode, sometimes switched to the diagnostic mode, producing unfiltered raw data in the original format. Alternatively, the diagnostics can be included in FPGA while data acquisition, which directly reports suspicious behavior.

Another counterargument against operating on packed information to save resources is no direct access by a human interpreter. However, it can be cheaply decoded for example by software and transformed into readable form. More important issue is susceptibility to data corruption - a single changed bit can make the entire compressed packet useless. This issue is usually neglected in standard protocols, where also a small change can result in large damage. It is usually resolved by including checksum and discarding entire corrupted packets, what can be also used here. In case data corruption is a serious problem, there can be applied a Forward Error Correction layer, which adds some redundancy to packets (increase their size), to be able to repair eventual damages.

The discussed data format comes from an universal time measurement device, therefore presents a significant overhead in order to support wide range of its possible configurations. We will present a concept of rearrangement and compression of that data assuming specific configuration and detector characteristics. Those characteristics define correlations between activity on tracking detector channels and the reference detector.

\section{Efficient data representation}
We will focus on the situation for a single event. The data sample has 48 channels, but we want to design a method which will be scalable for a larger number. Each channel is expected to produce rising-falling pairs of times, each of such pair corresponds to single digital pulse. The original time information consists of three times: $fine\_time\ \in \{0,\ldots,499]$ counts time in 10ps unit, up to 5ns. The $coarse\_time\ \in\{0,\ldots,2047\}$ counts time in 5ns units. Finally 28bit $epoch\_time$ counts time in unit of 5ns $\cdot$ 2048 = 10240ns.

Neglecting additional headers for clearance, the current coding consists of 2 types of 32bit data words:
\begin{itemize}
  \item main word consisting of 10 bits for $fine\_time$, 11 bits for $coarse\_time$, 7 bits for channel number, 1 bit for type of edge,
  \item words with mainly 28 bits of $epoch\_counter$.
\end{itemize}
First word type is used for each measurement. In case epoch has changed, the second word type is inserted additionally. In the data sample, the first type uses on average 700bits/event, the second 470 bits/event, what gives 1170 bits/event total.

We can see that there is already some wastefulness stemming from the requirement of working in 32 bit blocks: $fine\_time$ uses in fact 9 bits, channel number 6, there are some unused bits in such data blocks - for optimizations we need to use a more flexible structures of data.\\

Let us think of efficient data representation for this data.
\begin{itemize}
  \item Writing the channel number every time is a waste - instead, we could group values corresponding to a given channel,
  \item Writing absolute time is ineffective - we could write relative times instead (differences), which should be much smaller and can have some characteristic probability distribution, especially for the difference between rising and falling time and correlations between channels.
\end{itemize}
Let us introduce $time$ value for a given event:
$$ time=fine\_time + 500(coarse\_time + 2048 epoch\_time) - ref$$
where $ref$ is the reference time value (absolute time) for a given event (e.g. from a reference detector) - it might be required to be stored  ($\approx$ 3 bytes per event). As events are usually treated as independent, in many situations this value could be stored using a lower accuracy or completely neglected. A universal way to choose the $ref$ is to take the minimal of all times in this event. This way we can assume here that $time$ is non-negative, takes zero value once per event, and in the analysed data sample it always fits 32 bits. A different choice of $ref$ can be easily handled (generally can allow for negative $time$), and will be omitted in this analysis.

\begin{figure}[t!]        
    \centering
        \includegraphics[width=8.5cm]{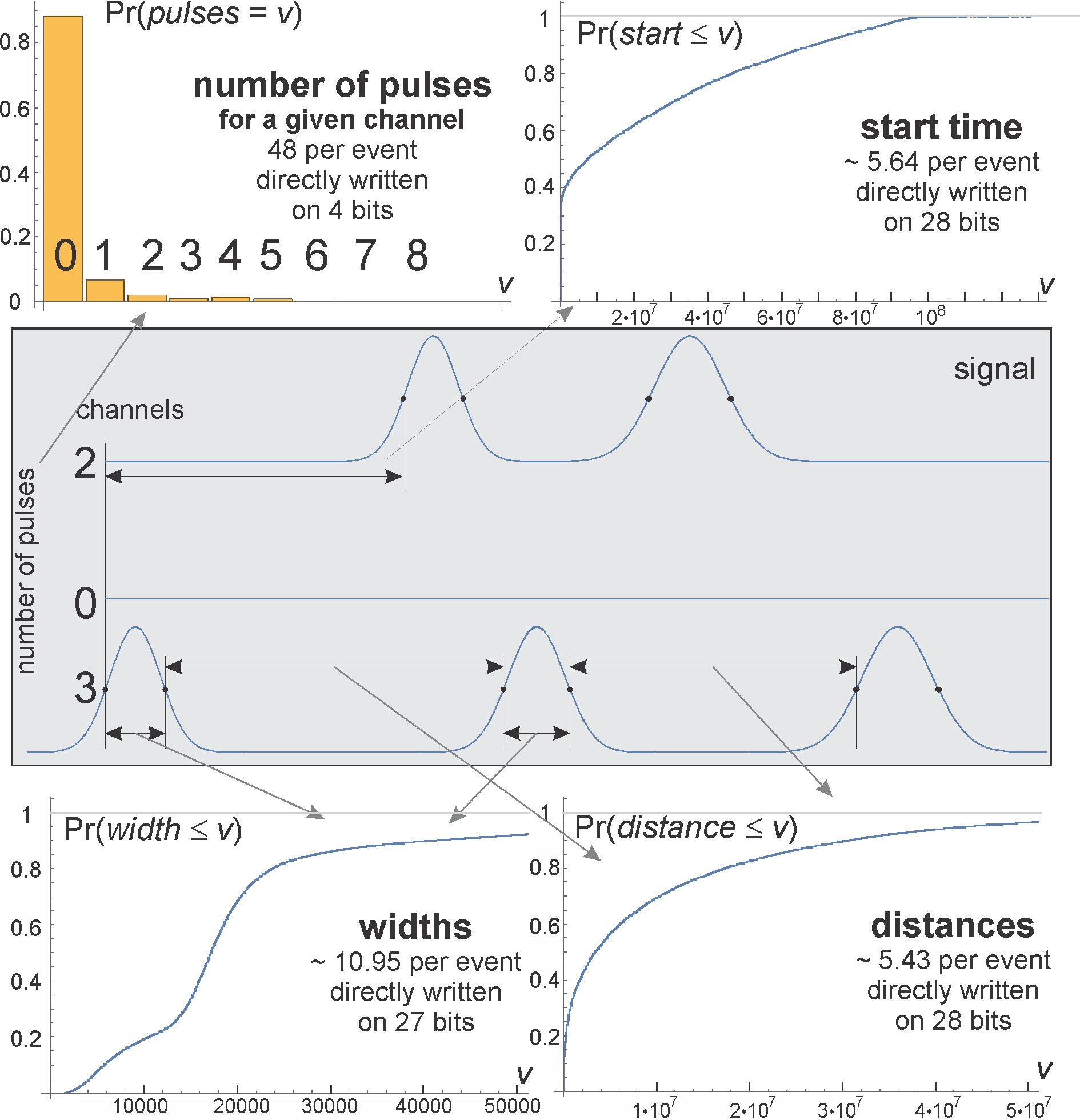}\
\begin{center}
        \caption{The proposed representation of time data from a single event and empirical cumulative distribution function (CDF) of values to store obtained from the data sample. For each channel we write the number of measured digital pulses ($pulses$). If it is nonzero, we also write $start$ as the difference between its first signal and the reference time (it is zero once per event), and then a series of $width$ then $distance$ until encoding all pulses. Four graphs show statistics for the analyzed data sample: the graph for \emph{pulses} shows probabilities of successive values, the graphs for \emph{start}, \emph{width} and \emph{distance} show empirical CDF, which was obtained by sorting the observed values.}
\end{center}
\label{general}
\end{figure}

Instead of writing the channel number, let us group all signals corresponding to a given channel. They are expected to appear in pairs corresponding to a single pulse: rising time, then falling time. Then a given channel waits some time for another digital pulse. So for each channel we need to store the moment of its first signal ($start$) which should be rising edge, then $width$ as the time difference until the signal of falling edge. If there are more pulses for this channel, we should store $distance$ as time difference to the next rising moment, then analogously $width$ and so on. Additionally, we need to indicated the number of pulses for each channel. There are different ways to realize it, for simplicity let us imagine that for each channel we write $pulses$ as the number of pulses for this channel. Finally, the situation looks like if Fig. \ref{general}.

We could now directly write such successive values using $\lfloor log_2 (max)\rfloor+1$ bits, where $max$ is the maximal value for a given type, getting correspondingly 4, 28, 27, 28 bits/value for $pulses$, $start$, $width$, $distance$ types of value. Multiplying them by average number of such values per event in our data sample and summing, we get on average 798 bits/event, what is 68\% of the original 1170 bits/event. The disadvantage is no longer operating on 32 bit blocks, but on 4, 28, 27, 28 length bit blocks occurring in a complex, history dependent order. However, decoding such data is a simple task for a computer.

Observe that by the way we have also saved the single bits characterizing type of edge, which is determined from the order here. However, this approach requires that data is already filtered - there is written only meaningful data: pairs of successive rising - falling times in our case. There could be written also data not fitting to such pattern, for example attached in the standard data format at the end of event. These exceptions, which will be usually discarded while analysis, are not fitting the pattern used for optimization, hence their storage is more costly than of the regular data.

Note that, while we focus on 48 channels, this solution can be naturally scaled to a larger number of channels, what will be required in the real application. It also perfectly fits the current architecture, where a few peripheral FPGA units, with less resources process data from some separate subsets of channels, and send it to some powerful central FPGA unit, which combines them into data packets of variable length. The peripheral FPGA units obtain time signals which are already sorted in time - it can directly calculate time differences ($width$ and $distance$), eventually perform some filtering: discard meaningless signals (e.g. two successive rising edges, extremely long $width$), and maybe also apply some data compression techniques. \\

Figure \ref{general} contains empirical probability distribution for $pulses$ values for our data sample, and empirical cumulative distribution functions (CDF) for $start$, $width$ and $distance$. Specifically, denoting $val[i]:\ i=1\ldots n$ as sorted values for a given type, such empirical CDF graph is plot of $(v[i],i/n)$ points. Directly storing numbers is optimal if they have uniform probability distribution (CDF is line for values in $(0,1)$), what is not the case here as we can see from the graphs. Entropy coder can be used to nearly optimally encode symbols from a general probability distribution, at cost of variable length coding: lengths of stored values are no longer fixed (like 4, 28, 27, 28), but depend on probabilities.
\section{Entropy coding: prefix codes and \lowercase{t}ANS}
Let us now focus on encoding values obtaining a small number of possibilities like $pulses$ in our case. In the next section we will look at storing larger numerical values. Data sample results in the following frequencies for the number of pulses per channel (starting with 0):
$$ (0.8825,\ 0.06591,\ 0.01948,\ 0.009375,\ 0.01503,\ $$ $$0.00653,\ 0.00101,\ 0.00013,\ 2\cdot 10^{-6}) $$
These are 9 possibilities, so directly storing them would require 4 bits/value. We could use a base 9 numeral system to store a sequence of 9 possibilities using asymptotically $\log_2(9)\approx 3.17$ bits/value.
\subsection{Prefix codes: Exp-Golomb and Huffman}
These coding options are optimal for uniform probability distributions among possibilities, while here we almost certainly will get $pulses=0$ value, so we should store this value using a smaller number of bits in our case. There is a simple, so called Exp-Golomb code (\cite{Golomb}) which is effective in such case of quickly decreasing probabilities for small natural numbers. Specifically, to encode a natural number $x$, we first write $\lfloor\log_2(x+1)\rfloor$ times "0" bit, then the entire binary expansion of $x+1$. For example we get $0\to 1,\ 1\to 010,\ 2\to 011,\ 3\to 00100,\ 4\to 00101$ and so on. Generally, value $x$ corresponds to length $2\lfloor\log_2(x+1)\rfloor+1$ bit sequence. Multiplying these lengths by probabilities and summing over all possibilities, we get the expected number of used bits/value, which is 1.30 in our $pulses$ case, what is 3 times better than the original 4 bits/value. However, as we can see in Fig. \ref{table}, Exp-Golomb coding is completely inefficient for our larger numerical values.

For a variable length coding like Exp-Golomb (the number of used bits depends on probability of symbol), we require prefix code condition: that bit sequence for a symbol/value is not a prefix of bit sequence for another value. Thanks of that, we can decode in unique way: the decoder always knows how many bits to use.

While Exp-Golomb coding does not take the actual probability distribution into consideration, Huffman coding (\cite{HC}) finds the optimal prefix code for a given distribution. It is done by repeating: retrieve two least probable symbols, group them into a new symbol with probability being the sum of the two probabilities, and put this new symbol into the alphabet. Each such grouping corresponds to a node in a binary tree, so finally we get a tree with symbols as its leafs. Now path from root to a given symbol corresponds to its bit sequence by translating left/right edges as 0/1 bits.

Finally, we get the following bit sequences for our $pulses$ values:
$$ (0,10,110,11110,1110,111110,1111110,11111110,111111110)$$
what in this case is very close to so called unary coding: for a value $x$ write "1" bit $x$ times, then write "0". This time we get a bit better: 1.23 bits/value.

We should also leave an option to handle exceptions, like $pulses>8$ possibilities. Assuming that they are very rare, despite the fact that such a single value is more costly, they should have practically negligible impact on average size, hence we are omitting them in this analysis.
\subsection{Shannon limit and \lowercase{t}ANS entropy coding}
Shannon entropy (\cite{Shannon}) is the theoretical limit for encoding a sequence from $\{p_s\}$ probability distribution - we asymptotically need on average:
$$ H= \sum_s p_s \log_2(1/p_s)\quad\textrm{bits/symbol}$$
in the case of $pulses$ values, it is 0.74 bits/value, what is much less than 1.23 for Huffman coding. Prefix codes operate on complete bits: have to use at least 1 bit/symbol. Generally symbol/event of probability $p$ carries $\log_2(1/p)$ bits of information, what can be much less than 1 bit/symbol. For example $pulses=0$ carries $\log_2(1/0.8825)\approx 0.18$ bits of information in our case.\\

There are also "accurate" entropy coders like arithmetic coding or range coding (\cite{ari,ran}), which allow to effectively approach the Shannon entropy limit by including the fractional numbers of bits. However, they require arithmetic multiplication, which is a costly resource, especially from the point of FPGA. There was recently introduced a more effective approach (Asymmetric Numeral Systems \cite{last,pcs2015}), which has already replaced Huffman and range coding in a few compressors to improve performance, like zhuff \cite{zhuff}, lzturbo \cite{lzturbo}, LZA \cite{LZA} or ZSTD \cite{ZSTD}. Its tabled variant (tANS) allows to approach Shannon entropy for a large alphabet without using multiplication, requiring a few kilobyte coding tables for 256 size alphabet. Software decoding can process  $\approx$ 500MB/s/core for a modern CPU, encoding $\approx 350$MB/s/core (\cite{fse}). In contrast, zlib implementation of Huffman coding (\cite{zlib}) has similar encoding speed, but only $\approx 300$MB/s/core decoding speed (and suboptimal compression ratio). tANS was also found suitable for FPGA implementations~\cite{ansfpga}.

\begin{figure}[t!]        
\centering
       \includegraphics[width=8.5cm]{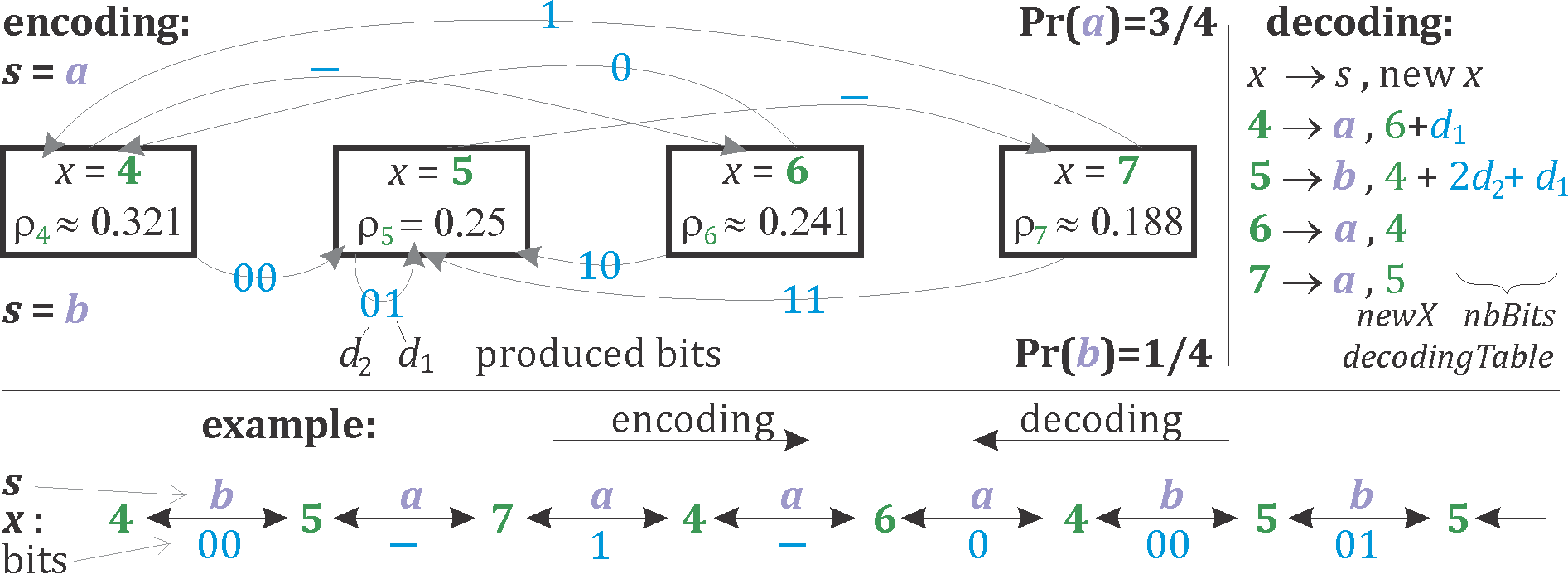}\
\begin{center}
        \caption{Example of 4 state tANS (top) and its application for stream coding (bottom) for two symbol alphabet of $\Pr(a)=3/4,\ \Pr(b)=1/4$ probabilities. State/buffer $x$ contains $\lg(x)\in [2,3)$ bits of information.
        Symbol $b$ carries 2 bits of information, while $a$ carries less than 1 - its information is gathered in $x$ until accumulating to a complete bit of information. The $\rho_x$ are probabilities of visiting state $x$ assuming i.i.d. input source.}        
\end{center}
\label{autom}
\end{figure}

We will now briefly present direct application of tANS method, more details can be found for example in \cite{pcs2015}. In this method we build a $L=2^R$ $(R\in \mathbb{N})$ state automaton dedicated for a given probability distribution, like depicted in Fig. \ref{autom} for $L=4$ states and $\Pr(a)=3/4,\ \Pr(b)=1/4$ probability distribution. Top-left part of this figure shows encoding step for symbol $a$ (upper) and $b$ (lower): for every symbol there is a set of rules for changing the state and eventually producing some bits (blue numbers on arrows). The top-right part of this figure presents decoding in this example: every state determines $symbol$ to decode, new state is $newX$ plus $nbBits$ of bits from the data stream, where $newX$ and $nbBits$ are also determined by the state.

While encoding we start from some chosen initial state ($x=4$ in the example), then encode successive symbols ("$baaaabb$"), leading to a bit sequence ("$00100001$") and a final state ($x=5$). Decoder needs to know this final state to start with, it process the bit sequence in backward order, producing symbol sequence in backward order. The inconvenience of backward decoding is usually resolved by encoding in backward direction in data frames of size of kilobytes, then decoding is straightforward. For FPGA encoding we can encode in forward direction and then software decode in backward direction instead. The cost of storing the final state once per data frame can be compensated by encoding some information in the initial state of encoder.

While prefix codes operated on integer number of bits, approximating probabilities by powers of $1/2$, tANS handles fractional number of bits thanks to the state $x\in \{L,\dots, 2L-1\}$ acting as a buffer containing $\log_2(x)\in [R,R+1)$ bits of information. This buffer gathers information and produces accumulated complete bits of information when needed.

In our example $x$ contains $\log_2(x)\in[2,3)$ bits of information. Symbol $b$ contains $\log_2(2)=2$ bits of information, and so we see that it always produces 2 bits of information here. Symbol $a$ contains $\log_2(4/3)<1$ bits of information, and so it sometimes produces one bit (from $x=6,7$), sometimes zero bits (from $x=4,5$) only increasing the state (accumulating information in the buffer) - symbol $a$ produces on average less than 1 bits/symbol.

To quantitatively evaluate performance of such entropy coder, we need first to find the probability distribution of visiting successive states: $\rho_x$ as the stationary probability distribution for such a random walk among states: assuming corresponding i.i.d. input source of symbols. Finally the automaton from Fig. \ref{autom} produces on average $H'\approx (0.241+0.188)\cdot 3/4\cdot 1 + 1\cdot 1/4 \cdot 2\approx H+0.01$ bits/symbol, where $H$ is the minimum: Shannon entropy. In other words, the inaccuracy of this entropy coder costs us using $\Delta H=H'-H \approx 0.01$ more bits/symbol than the optimum. It can be reduced by using more states, e.g. $L=8$ state automaton would give $\Delta H\approx 0.0018$ bits/symbol for the $\Pr(a)=3/4,\ \Pr(b)=1/4$ case. Generally, $\delta H$ drops approximately with square of the number of used states ($L$) and grows with square of alphabet size ($m$). So in practical applications there is chosen a fixed $L/m$ proportion, for example as 8 in FSE (\cite{fse}).

In a general case, above decoding/encoding steps, which are the critical loops, can be optimized to a compact form:

\begin{algorithm}[htbp]
\footnotesize{
\caption{tANS decoding step, $X=x-L\in\{0,\ldots,L-1\}$}
\label{dec0}
\begin{algorithmic}
\STATE $t = decodingTable[X]$  \qquad\quad  \COMMENT{ $X\in \{0,..,L-1\}$ is current state }
\STATE useSymbol($t.symbol$)    \qquad\qquad   \COMMENT{ use or store decoded symbol }
\STATE $X = t.newX + $readBits$(t.nbBits)$   \qquad\qquad  \COMMENT{ state transition }
\end{algorithmic}
}
\end{algorithm}

\begin{spacing}{0.5}
\begin{algorithm}[htbp]
\footnotesize{
\caption{tANS encoding step for symbol $s$ and state $x=X+L$}
\label{enc0}
\begin{algorithmic}
\STATE $nbBits = (x + nb[s]) >> r$    \quad\qquad \COMMENT{$r=R+1,\ 2^r = 2L$}
\STATE useBits$(x, nbBits)$ \qquad \qquad\COMMENT {use $nbBits$ of the youngest bits of $x$}
\STATE $x = encodingTable[start[s] + (x >> nbBits)]$
\end{algorithmic}
}
\end{algorithm}
\end{spacing}

Where $decodingTable$ for our example is: $symbol$ are correspondingly $\{a,b,a,a\}$, $nbBits$: $\{1,2,0,0\}$ and $newX$ are $\{6,4,4,5\}$. For encoding in our example we can choose $nb[a]=2,\ nb[b]=12,\ start[a]=-3,\ start[b]=2,\ encodingTable[0..3]=\{4,6,7,5\}$.\\

We will now present algorithms to produce such automaton for a general alphabet and probability distribution $\{p_s\}_{s=1..m}$ by generating the tables used in above encoding/decoding steps. Assume that $L=2^R$ and $0<L_s \approx L\cdot p_s$ approximate the probability distribution of symbols, such that $\sum_s L_s=L$. Now we need to choose a symbol spread function: $symbol[X]:\ X\in\{0,\ldots,L-1\}\to\{1,\ldots,m\}$, such that symbol $s$ appears $L_s$ times there: $L_s = \{X:\ symbol[X]=s\}$. This function defines the coding. In our example the symbol spread is correspondingly $\{a,b,a,a\}$, $L_a=3,\ L_b=1$.

The optimal choice of $symbol[X]$ symbol spread function is a complicated topic, we present only a fast simple way to spread symbols in a pseudorandom which is used in FSE and usually provide excellent performance. More sophisticated methods can offer a small improvements of $\Delta H$ - many of them can be found and tested in \cite{toolkit}.

\begin{spacing}{0.2}
\begin{algorithm}[htbp]
\footnotesize{
\caption{An example of fast symbol spread function \cite{fse}}
\label{spread2}
\begin{algorithmic}
\STATE $X=0$; $step=5/8 L +3$ \quad \COMMENT{ some initial position and choice of step}
\FOR {$s=0$ to $m-1$}
\FOR {$i=1$ to $L_s$}
\STATE $symbol[X]=s$; $X = $mod$(X + step, L)$
\ENDFOR
\ENDFOR
\end{algorithmic}
}
\end{algorithm}
\end{spacing}

After choosing the symbol spread $symbol[X]$, Method~\ref{gen} generates the $decodingTable$ for decoding step from Method~\ref{dec0}. For efficient memory handling while encoding step, the encoding table can be stored in one dimensional form $encodingTable[x + start[s]] \in I$ for $x\in \{L_s,\ldots,2L_s-1\}$, where $start[s]=- L_s + \sum_{s'<s}L_{s'} $.
To encode symbol $s$ from state $x$, we need first to transfer $k[s]-1$ or $k[s]$ bits, where $k[s] = \lceil lg(L/L_s) \rceil$. The smallest $x$ for $k[s]$ bits is $bound[s] = L_s \cdot 2^{k[s]} \in \{L,\ldots,2L-1\}$. Finally, preparation and encoding step are written as Methods~\ref{encprep} and \ref{enc0} correspondingly.

\begin{algorithm}[htbp]
\footnotesize{
\caption{Preparation for tANS decoding, $L=2^R$}
\label{gen}
\begin{algorithmic}
\REQUIRE $next[s] = L_s$ \qquad\COMMENT{number of next appearance of symbol $s$}
\FOR {$X=0$ to $L-1$}
\STATE $t.symbol=symbol[X]$ \qquad\COMMENT{ symbol is from spread function }
\STATE $x=next[t.symbol]++$ \qquad \COMMENT{ $D(X+L)=(symbol,x)$ }
\STATE $t.nbBits = R -\lfloor \lg(x)\rfloor$ \qquad \COMMENT{ number of bits}
\STATE $t.newX = (x << t.nbBits)-L$\qquad\COMMENT{ properly shift $x$ }
\STATE $decodingTable[X]=t$
\ENDFOR
\end{algorithmic}
}
\end{algorithm}

\begin{algorithm}[htbp]
\footnotesize{
\caption{Preparation for tANS encoding, $L=2^R$, $r=R+1$}
\label{encprep}
\begin{algorithmic}
\REQUIRE $k[s] = R-\lfloor \lg(L_s) \rfloor$  \qquad\qquad\quad \COMMENT{$nbBits=k[s]$ or $k[s]-1$}
\REQUIRE $nb[s] = (k[s] << r)-(L_s << k[s])$
\REQUIRE $start[s]= - L_s + \sum_{s'<s}L_{s'}$
\REQUIRE $next[s] = L_s$
\FOR {$x=L$ to $2L-1$}
\STATE $s=symbol[x-L]$
\STATE $encodingTable[start[s] + (next[s]++)] = x$;
\ENDFOR
\end{algorithmic}
}
\end{algorithm}

\section{Simple and adaptive binning}
Our numerical values $start$, $width$, $distance$ are rather too large to be directly written by an entropy coder. However, their least significant bits have often nearly uniform probability distribution, so directly writing them can be already effective. In contrast, their most significant bits may have very nonuniform probability distribution - we can use entropy coder to optimize their storage cost.

\begin{figure}[t!]        
    \centering
        \includegraphics[width=8.5cm]{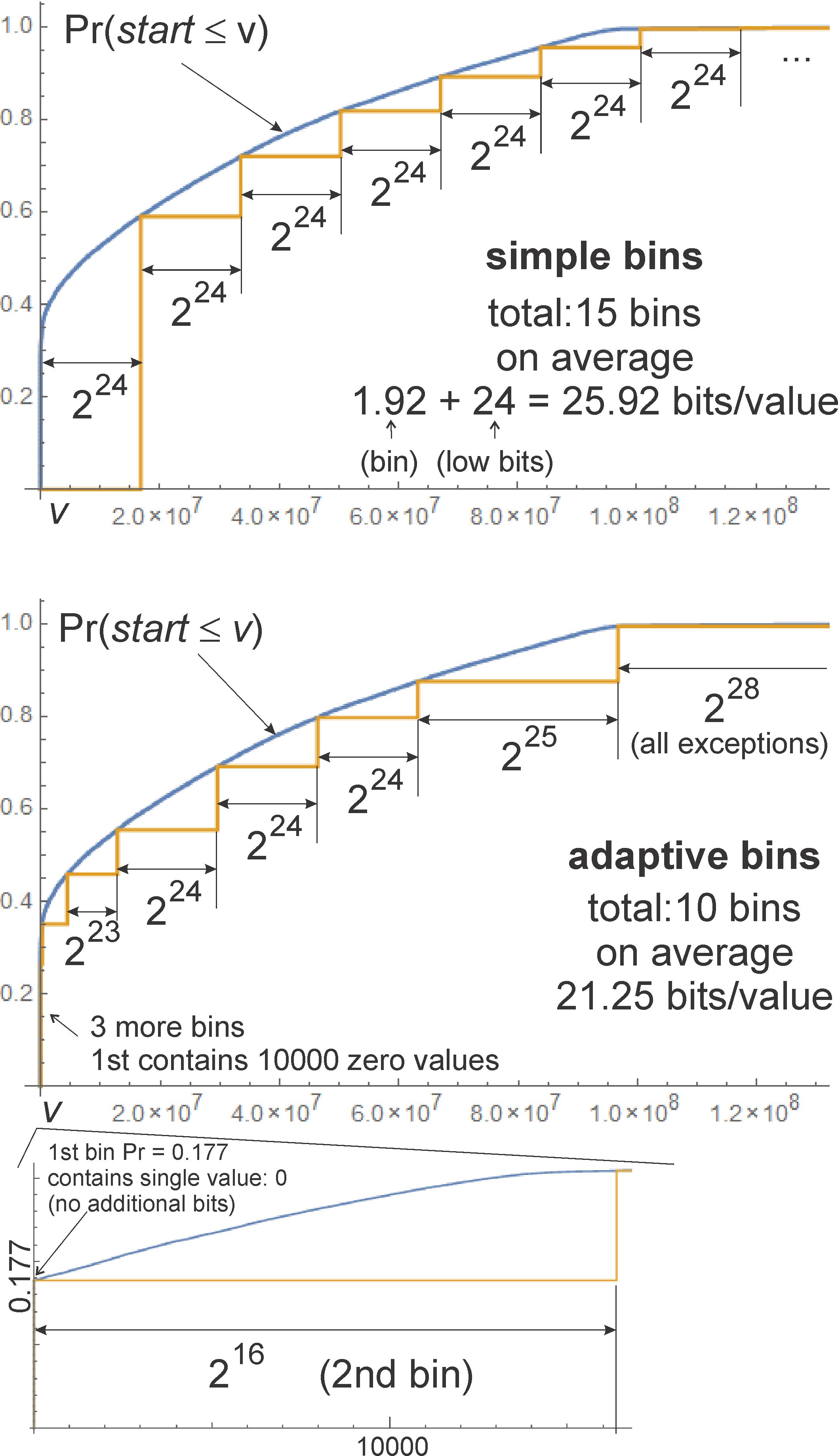}\
\begin{center}
        \caption{Directly writing the \emph{start} values would cost 28bits/value. In the above 24 bit simple binning we use entropy coder to encode the most significant 4 of these bits, what costs on average 1.92 bits/value and 24 bits to encode the remaining least significant bits. The lower part shows adaptive banning (and magnification for small $start$ values): we divide the range into variable-width bins, optimized to have approximately uniform probability distribution of the least significant bits (CDF is approximately linear there). The choice of bin costs on average 3.15 bits/symbol and 18.09 bits/value is the average cost of storing the remaining least significant bits. 10000 out of 56385 $start$ values were zero, hence the first bin of probability 17.7\% just produces the zero value without using the further bits. For 20bit simple binning we get 24.85 bits/value for 237 bins, for 168 adaptive bins we get 21.06 bits/value.}
\end{center}
\label{binning}
\end{figure}

The simplest approach is to write some number of the most significant bits using entropy coder and then directly write the remaining bits, what will be referred as simple binning. The top part of Fig. \ref{binning} shows its example for our $start$ values - split into 15 bins for the most significant 4 bits, the remaining 24 bits should be directly written. The probability of a bin can be obtained as a percentage of values using this bin.

The assumption of nearly uniform probability distribution is often violated, especially for small values. For example most of values in the first bin of our $start$ example will be zero - storing 24 low bits is a waste here. Hence, adjusting bin sizes to a given case can be beneficial, especially when
\begin{itemize}
  \item there are special cases like $start=0$ in our example,
  \item for quickly decaying probability distributions like Gaussian (see $20.1 \to 15.5$ improvement for $width$ in Fig. \ref{table}),
  \item when we want to handle exceptions we can use large bins at both ends, what would require using a large number of simple bins, making it costly from the perspective of entropy coding.
\end{itemize}

For adaptive binning we need two tables: $binStart[i]$ as the minimal value for $i$-th bin, and $binWidth[i]$ as the number of the least significant bits required to choose a value inside this bin. The next bin starts at the successive position: $startBin[i+1]=startBin[i]+2^{binWidth[i]}$.

Instead of the bin number, entropy decoder can directly return such two numbers: $binStart$, $binWidth$ and decoded value is
$$v=binStart + \textrm{readBits}(binWidth)$$
Encoding is a bit more complicated as we need to determine the bin. One possibility is to use a table which takes some number of the most significant bits of value (e.g. 8) and directly returns the bin number if it is unique. Otherwise, it could point to analogous another table for successive e.g. 8 bits, and so on if required. Special cases like our $start=0$ can be handled by a separate condition.

There have remained a question of choosing adaptive binning. The optimal choice seems to be a complicated problem. The presented evaluation used a simple heuristics and manipulation of a $minVal$ parameter: the minimal number of values per bin. Specifically, after sorting values, we construct successive bins by increasing $binWidth$ until exceeding the $minValn$ number of values inside this bin or reaching the end. This simple algorithm quickly approaches some asymptotic average cost, suggesting it is sufficient for practical applications.
\section{Separate channel distributions and further optimizations}
\begin{figure}[b!]        
    \centering
        \includegraphics[width=8.5cm]{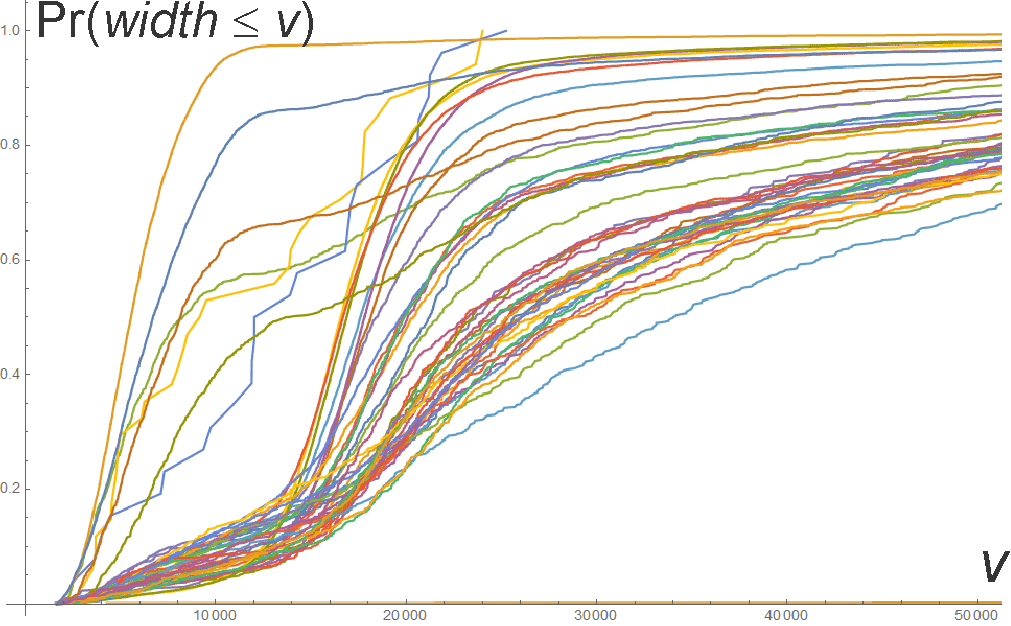}\
\begin{center}
        \caption{Empirical CDF for widths for separate channels - each color corresponds to a different channel. Using a single entropy coder for all of them we need on average $\approx$15.5 bits/value, while using a separate optimized for each of them reduces this value to $\approx$14.9 bits/value.}
\end{center}
\label{separate}
\end{figure}

In the previous sections we were assuming that all channels use the same probability distribution, the statistics were obtained by putting all values of given type into one box. Looking separately at each channel: dividing the data into 48 boxes, we can see that data from separate channels seems to be governed by separate statistical rules, as a result of geometry of the experiment. For example Fig. \ref{separate} shows $width$ empirical CDFs for separate channels - nearly all of them resemble CDF for Gaussian distribution, but of different ones.

These differences are a consequence of geometry of the experiment. We could exploit them by having separate sets of coding tables, each one optimized for a given channel, leading to approximately 4\% income for our sample data. Such coding would become more memory demanding, so such full separation is rather restricted to software compressors. However, one could consider intermediate solution to get intermediate improvements, for example classifying channels into a few classes, and use a separate set of coding tables for every class. Additionally, in real experiment an FPGA unit should process data from detectors which are close to each other and so should have similar statistical behavior - there could be used coding tables optimized for a given FPGA.

There have still remained place for other optimizations. For example we know that exactly once per event there will be $start$=0 value. Additionally, it is more likely to happen for a channel with large $pulses$. Approximately 2 bits/event could be saved by pointing the $start$=0 channel and encode only nonzero $start$ values for the remaining channels.

Finally, while we have assumed that signals from separate channels are independent, there should be some hidden correlations which exploitation could lead to further essential savings. For example concentration of hits in some channels may suggest increased activity in other channels. An idealized compressor should first classify type of event and estimate its parameters, then use them to estimate probability distributions for channel activations, to be used for optimized data encoding. This topic requires further research.
\section{Conclusions}
Efficient encoding of data from physics experiments can be one of tools to reduce required resources: data storage and transmission lines. It can also improve the processing speed as decoding is usually much faster than reading from HDD. The basic suggestions for designing such data acquisition systems and protocols are:
\begin{itemize}
  \item Separate diagnostic runs from the proper data acquisition - when early filtering can be used to transmit and store only the meaningful information - which will be actually used in the analysis. FPGA usually used for acquisition may have unused potential to include some initial filtering and optimized encoding.
  \item Instead of writing labels (like channel) for each data block, try to group information having the same label.
  \item Instead of writing absolute values like time, relative values (differences) have usually much smaller and more predictable values, often come from some characteristic distributions like exponential or Gaussian.
  \item Use entropy coder, especially for probability distributions which are far from uniform.
  \item Try to exploit correlations, for example predict values using previous ones and encode the difference.
\end{itemize}

\footnotesize
\bibliographystyle{IEEEtran}
\bibliography{ref}
\end{document}